\magnification=1200

\def\gl{\mathrel{\raise1ex\hbox{$>$\kern-.75em\lower1ex\hbox{$<$}}}}
\def\lg{\mathrel{\raise1ex\hbox{$<$\kern-.75em\lower1ex\hbox{$>$}}}}
\def\gtwid{\mathrel{\raise.3ex\hbox{$>$\kern-.75em\lower1ex\hbox{$\sim$}}}}
\def\ltwid{\mathrel{\raise.3ex\hbox{$<$\kern-.75em\lower1ex\hbox{$\sim$}}}}
\def\sqr#1#2{{\vcenter{\hrule height.#2pt
      \hbox{\vrule width.#2pt height#1pt \kern#1pt
         \vrule width.#2pt}
      \hrule height.#2pt}}}

\overfullrule=0pt


\def\eg{\hbox{{\it e.\ g.}}}\def\ie{\hbox{{\it i.\ e.}}}


\def\leaderfill{\leaders\hbox to 1em{\hss.\hss}\hfill}


\def\CC{\hbox{{$\cal C$}}} 
\def\CD{\hbox{{$\cal D$}}}

\def\CR{\hbox{{$\cal R$}}}

\def\ref#1{~[#1]}
\newcount\eqnum \eqnum=0  
\newcount\eqnA\eqnA=0\newcount\eqnB\eqnB=0\newcount
\eqnC\eqnC=0\newcount\eqnD\eqnD=0
\def\eqnoi{\global\advance\eqnum by 1\eqno(\the\eqnum)}
\def\eqnai{\global\advance\eqnum by 1\eqno(\the\eqnum{a})}
\def\eqnbi{\eqno(\the\eqnum{b})}
\def\eqnci{\eqno(\the\eqnum{c})}
\def\eqnoA{\global\advance\eqnA by 1\eqno({\rm A}\the\eqnA)}
\def\eqnoB{\global\advance\eqnB by 1\eqno({\rm B}\the\eqnB)}
\def\eqnoC{\global\advance\eqnC by 1\eqno({\rm C}\the\eqnC)}
\def\eqnoD{\global\advance\eqnD by 1\eqno({\rm D}\the\eqnD)}
\def\back#1{{\advance\eqnum by-#1 Eq.~(\the\eqnum)}}
\def\backa#1{{\advance\eqnum by-#1 Eq.~(\the\eqnum(a))}}
\def\backb#1{{\advance\eqnum by-#1 Eq.~(\the\eqnum(b))}}
\def\backc#1{{\advance\eqnum by-#1 Eq.~(\the\eqnum(c))}}
\def\backs#1{{\advance\eqnum by-#1 Eqs.~(\the\eqnum)}}
\def\backn#1{{\advance\eqnum by-#1 (\the\eqnum)}}
\def\backA#1{{\advance\eqnA by-#1 Eq.~(A\the\eqnA)}}
\def\backB#1{{\advance\eqnB by-#1 Eq.~(B\the\eqnB)}}
\def\backC#1{{\advance\eqnC by-#1 Eq.~(C\the\eqnC)}}
\def\backD#1{{\advance\eqnD by-#1 Eq.~(D\the\eqnD)}}
\def\last{{Eq.~(\the\eqnum)}}                    
\def\lasta{{Eq.~(\the\eqnum{a})}}                
\def\lastb{{Eq.~(\the\eqnum{b})}}                
\def\lastc{{Eq.~(\the\eqnum{c})}}                
\def\lasts{{Eqs.~(\the\eqnum)}}                  
\def\lastn{{(\the\eqnum)}}                       
\def\backn#1{{\advance\eqnum by-#1 (\the\eqnum)}}
\def\lastA{{Eq.~(A\the\eqnA)}}\def\lastB{{Eq.~(B\the\eqnB)}}
\def\lastC{{Eq.~(C\the\eqnC)}}\def\lastD{{Eq.~(D\the\eqnD)}}
\newcount\refnum\refnum=0  
\def\refi{\smallskip\global\advance\refnum by 1\item{\the\refnum.}}

\newcount\rfignum\rfignum=0 
\def\rfigi{\medskip\global\advance\rfignum by 1\item{Figure \the\rfignum.}}

\newcount\fignum\fignum=0  
\def\figi{\global\advance\fignum by 1 Fig.~\the\fignum}

\newcount\rtabnum\rtabnum=0 
\def\rtabi{\medskip\global\advance\rtabnum by 1\item{Table \the\rtabnum.}}

\newcount\tabnum\tabnum=0  
\def\tabi{\global\advance\tabnum by 1 Table~\the\tabnum}

\newcount\secnum\secnum=0 
\def\chap#1{\global\advance\secnum by 1
\bigskip\centerline{\bf{\the\secnum}. #1}\smallskip\noindent}

\def\lsubsec#1{\global\advance\secnum by 1
\smallskip{\bf\noindent{\the\secnum. #1}}}
\def\tlsubsec#1{\global\advance\secnum by 1
{\bf\noindent{\the\secnum. #1}}}

\newcount\nlet\nlet=0
\def\numblet{\relax \global\advance\nlet by 1
\ifcase \nlet \ \or a\or b\or c\or d\or e\or
f\or g\or h\or i\or j\or k\or l\or m\or n\or o\or p
\or q \or r \or s \or t \or u \or v \or w \or x \or y
\or z \else .\nlet \fi}

\newcount\rslet\rslet=0
\def\romslet{\relax \global\advance\rslet by 1
\ifcase \rslet \ \or i\or ii\or iii\or iv\or v\or
vi\or vii\or viii\or ix\or x\or xi\or xii \else .\rslet \fi}

\newcount\rlet\rlet=0
\def\romlet{\relax \global\advance\rlet by 1
\ifcase \rlet \ \or I\or II\or III\or IV\or V\or
VI\or VII\or VIII\or IX\or X\or XI\or XII\or XIII\or XIV\or XV\or XVI
\or XVII \else .\rlet \fi}
\def\rchap#1{\bigskip\centerline{\bf{\romlet}. #1}\smallskip\noindent}

\def\pd#1#2{{\partial #1\over\partial #2}}      
\def\p2d#1#2{{\partial^2 #1\over\partial #2^2}} 

\def\td#1#2{{d #1\over d #2}}      
\def\t2d#1#2{{d^2 #1\over d #2^2}} 


\def\2kth{{$2k^{\rm th}$}}

\def\n-th{{$(n-1)^{\rm th}$}}

\def\N-th{{$(N-1)^{\rm th}$}}

\def\0th{$0^{\rm th}$}
\def\1st{$1^{\rm st}$}
\def\2nd{$2^{\rm nd}$}
\def\3rd{$3^{\rm rd}$}
\def\4th{$4^{\rm th}$}
\def\5th{$5^{\rm th}$}
\def\5th{$6^{\rm th}$}
\def\6th{$7^{\rm th}$}
\def\7th{$7^{\rm th}$}
\def\8th{$8^{\rm th}$}
\def\9th{$9^{\rm th}$}

\def\a{{\alpha}}
\def\b{{\beta}}
\def\g{\gamma}

\def\n{\nu}

\def\t{\tau}

\def\aam #1 #2 #3 {{\sl Adv.\ Appl.\ Mech.} {\bf #1}, #2 (#3)}
\def\aap #1 #2 #3 {{\sl Adv.\ Appl.\ Prob.} {\bf #1}, #2 (#3)}
\def\ac #1 #2 #3 {{\sl Adv.\ Catal.} {\bf #1}, #2 (#3)}
\def\aces #1 #2 #3 {{\sl Adv.\ Chem.\ Eng.\ Sci.} {\bf #1}, #2 (#3)}
\def\acp #1 #2 #3 {{\sl Adv.\ Chem.\ Phys.} {\bf #1}, #2 (#3)}
\def\ae #1 #2 #3 {{\sl Ann.\ Eugenics} {\bf #1}, #2 (#3)}
\def\ah #1 #2 #3 {{\sl Adv.\ Hydrosci.} {\bf #1}, #2 (#3)}
\def\aichej #1 #2 #3 {{\sl AICHE J.} {\bf #1}, #2 (#3)}
\def\aj #1 #2 #3 {{\sl Astrophys.\ J.} {\bf #1}, #2 (#3)}
\def\ajp #1 #2 #3 {{\sl Am.\ J.\ Phys.} {\bf #1}, #2 (#3)}
\def\ams #1 #2 #3 {{\sl Ann.\ Math.\ Stat.} {\bf #1}, #2 (#3)}
\def\ap #1 #2 #3 {{\sl Adv.\ Phys.} {\bf #1}, #2 (#3)}
\def\apa #1 #2 #3 {{\sl Appl.\ Phys.\ A} {\bf #1}, #2 (#3)}
\def\apc #1 #2 #3 {{\sl Appl.\ Catal.} {\bf #1}, #2 (#3)}
\def\annp #1 #2 #3 {{\sl Ann.\ Phys.\ (N.Y.)} {\bf #1}, #2 (#3)}
\def\annpr #1 #2 #3 {{\sl Ann.\ Probab.} {\bf #1}, #2 (#3)}
\def\arfm #1 #2 #3 {{\sl Ann.\ Rev.\ Fluid Mech.} {\bf #1}, #2 (#3)}
\def\arpc #1 #2 #3 {{\sl Ann.\ Rev.\ Phys.\ Chem.} {\bf #1}, #2 (#3)}
\def\astech #1 #2 #3 {{\sl Aerosol Sci.\ Tech.} {\bf #1}, #2 (#3)}
\def\bcdg #1 #2 #3 {{\sl Ber.\ Chem.\ Dtsch.\ Ges.} {\bf #1}, #2 (#3)}
\def\bio #1 #2 #3 {{\sl Biometrika} {\bf #1}, #2 (#3)}
\def\bj #1 #2 #3 {{\sl Biophys.\ J.} {\bf #1}, #2 (#3)}
\def\ces #1 #2 #3 {{\sl Chem.\ Engr.\ Sci.} {\bf #1}, #2 (#3)}
\def\cf #1 #2 #3 {{\sl Combust.\ and Flame} {\bf #1}, #2 (#3)}
\def\cmp #1 #2 #3 {{\sl Commun.\ Math.\ Phys.} {\bf #1}, #2 (#3)}
\def\cp #1 #2 #3 {{\sl Chem.\ Phys.} {\bf #1}, #2 (#3)}
\def\contp #1 #2 #3 {{\sl Contemp.\ Phys.} {\bf #1}, #2 (#3)}
\def\cpam #1 #2 #3 {{\sl Commun.\ Pure Appl.\ Math.} {\bf #1}, #2 (#3)}
\def\crev #1 #2 #3 {{\sl Catal.\ Rev.} {\bf #1}, #2 (#3)}
\def\crII #1 #2 #3 {{\sl C. R. Acad.\ Sci.\ Ser.\ II} {\bf #1}, #2 (#3)}
\def\eul #1 #2 #3 {{\sl Europhys.\ Lett.} {\bf #1}, #2 (#3)}
\def\ic #1 #2 #3 {{\sl Icarus} {\bf #1}, #2 (#3)}
\def\iec #1 #2 #3 {{\sl Ind.\ Eng.\ Chem.} {\bf #1}, #2 (#3)}
\def\ijf #1 #2 #3 {{\sl Int.\ J.\ of Fracture} {\bf #1}, #2 (#3)}
\def\ijmpA #1 #2 #3 {{\sl Int.\ J.\ Modern Phys. A} {\bf #1}, #2 (#3)}
\def\ijmpB #1 #2 #3 {{\sl Int.\ J.\ Modern Phys. B} {\bf #1}, #2 (#3)}
\def\ijss #1 #2 #3 {{\sl Int.\ J.\ Solids Structures} {\bf #1}, #2 (#3)}
\def\jam #1 #2 #3 {{\sl J.\ Appl.\ Mech.} {\bf #1}, #2 (#3)}
\def\jap #1 #2 #3 {{\sl J.\ Appl.\ Phys.} {\bf #1}, #2 (#3)}
\def\japr #1 #2 #3 {{\sl J.\ Appl.\ Prob.} {\bf #1}, #2 (#3)}
\def\jaes #1 #2 #3 {{\sl J.\ Aerosol.\ Sci.} {\bf #1}, #2 (#3)}
\def\jas #1 #2 #3 {{\sl J.\ Atmos.\ Sci.} {\bf #1}, #2 (#3)}
\def\jasa #1 #2 #3 {{\sl J. Acous.\ Soc.\ Amer.} {\bf #1}, #2 (#3)}
\def\jc #1 #2 #3 {{\sl J.\ Catal.} {\bf #1}, #2 (#3)}
\def\jce #1 #2 #3 {{\sl J.\ Chem.\ Educ.} {\bf #1}, #2 (#3)}
\def\jcis #1 #2 #3 {{\sl J.\ Colloid Interface Sci.} {\bf #1}, #2 (#3)}
\def\jcg #1 #2 #3 {{\sl J.\ Crystal Growth} {\bf #1}, #2 (#3)}
\def\jcp #1 #2 #3 {{\sl J.\ Chem.\ Phys.} {\bf #1}, #2 (#3)}
\def\jcompp #1 #2 #3 {{\sl J.\ Comp.\ Phys.} {\bf #1}, #2 (#3)}
\def\jdep #1 #2 #3 {{\sl J.\ de Physique I} {\bf #1}, #2 (#3)}
\def\jdepl #1 #2 #3 {{\sl J. de Physique Lett.} {\bf #1}, #2 (#3)}
\def\jetp #1 #2 #3 {{\sl Sov.\ Phys.\ JETP} {\bf #1}, #2 (#3)}
\def\jetpl #1 #2 #3 {{\sl Sov. Phys.\ JETP Letters} {\bf #1}, #2 (#3)}
\def\jes #1 #2 #3 {{\sl J. Electrochem.\ Soc.} {\bf #1}, #2 (#3)}
\def\jfi #1 #2 #3 {{\sl J. Franklin Inst.} {\bf #1}, #2 (#3)}
\def\jfm #1 #2 #3 {{\sl J. Fluid Mech.} {\bf #1}, #2 (#3)}
\def\jgr #1 #2 #3 {{\sl J.\ Geophys.\ Res.} {\bf #1}, #2 (#3)}
\def\jif #1 #2 #3 {{\sl J. Inst.\ Fuel} {\bf #1}, #2 (#3)}
\def\jmo #1 #2 #3 {{\sl J. Mod. Opt.} {\bf #1}, #2 (#3)}
\def\jmp #1 #2 #3 {{\sl J. Math. Phys.} {\bf #1}, #2 (#3)}
\def\jms #1 #2 #3 {{\sl J. Memb.\ Sci.} {\bf #1}, #2 (#3)}
\def\josaA #1 #2 #3 {{\sl J. Opt.\ Soc.\ Am.\ A} {\bf #1}, #2 (#3)}
\def\josaB #1 #2 #3 {{\sl J. Opt.\ Soc.\ Am.\ B } {\bf #1}, #2 (#3)}
\def\jpa #1 #2 #3 {{\sl J. Phys.\ A} {\bf #1}, #2 (#3)}
\def\jpc #1 #2 #3 {{\sl J. Phys.\ C} {\bf #1}, #2 (#3)}
\def\jpd #1 #2 #3 {{\sl J. Phys.\ D} {\bf #1}, #2 (#3)}
\def\jpchem #1 #2 #3 {{\sl J. Phys.\ Chem.} {\bf #1}, #2 (#3)}
\def\jps #1 #2 #3 {{\sl J. Polymer.\ Sci.} {\bf #1}, #2 (#3)}
\def\jpsj #1 #2 #3 {{\sl J. Phys.\ Soc. Jpn.} {\bf #1}, #2 (#3)}
\def\jpso #1 #2 #3 {{\sl J. Power Sources} {\bf #1}, #2 (#3)}
\def\jsc #1 #2 #3 {{\sl J. Sci.\ Comp.} {\bf #1}, #2 (#3)}
\def\jsp #1 #2 #3 {{\sl J. Stat.\ Phys.} {\bf #1}, #2 (#3)}
\def\jtb #1 #2 #3 {{\sl J. Theor.\ Biol.} {\bf #1}, #2 (#3)}
\def\kc #1 #2 #3 {{\sl Kinet.\ Catal.\ (USSR)} {\bf #1}, #2 (#3)}
\def\macro #1 #2 #3 {{\sl Macromolecules} {\bf #1}, #2 (#3)}
\def\mclc #1 #2 #3 {{\sl Mol.\ Cryst.\ Liq.\ Cryst.} {\bf #1}, #2 (#3)}
\def\mubm #1 #2 #3 {{\sl Moscow Univ.\ Bull.\ Math.} {\bf #1}, #2 (#3)}
\def\nat #1 #2 #3 {{\sl Nature} {\bf #1}, #2 (#3)}
\def\npa #1 #2 #3 {{\sl Nucl.\ Phys.\ A} {\bf #1}, #2 (#3)}
\def\npb #1 #2 #3 {{\sl Nucl.\ Phys. B} {\bf #1}, #2 (#3)}
\def\PA #1 #2 #3 {{\sl PAGEOPH} {\bf #1}, #2 (#3)}
\def\pA #1 #2 #3 {{\sl Physica A} {\bf #1}, #2 (#3)}
\def\pB #1 #2 #3 {{\sl Physica B} {\bf #1}, #2 (#3)}
\def\pBC #1 #2 #3 {{\sl Physica B \& C} {\bf #1}, #2 (#3)}
\def\pD #1 #2 #3 {{\sl Physica D} {\bf #1}, #2 (#3)}
\def\pams #1 #2 #3 {{\sl Proc.\ Am.\ Math.\ Soc.} {\bf #1}, #2 (#3)}
\def\pcps #1 #2 #3 {{\sl Proc.\ Camb.\ Philos.\ Soc.} {\bf #1}, #2 (#3)}
\def\pfA #1 #2 #3 {{\sl Phys.\ Fluids A} {\bf #1}, #2 (#3)}
\def\pla #1 #2 #3 {{\sl Phys.\ Lett. A} {\bf #1}, #2 (#3)}
\def\plb #1 #2 #3 {{\sl Phys.\ Lett. B} {\bf #1}, #2 (#3)}
\def\pmB #1 #2 #3 {{\sl Philos.\ Mag. B} {\bf #1}, #2 (#3)}
\def\pnas #1 #2 #3 {{\sl Proc.\ Natl.\ Acad.\ Sci.} {\bf #1}, #2 (#3)}
\def\pr #1 #2 #3 {{\sl Phys.\ Rev.} {\bf #1}, #2 (#3)}
\def\pra #1 #2 #3 {{\sl Phys.\ Rev.\ A} {\bf #1}, #2 (#3)}
\def\prb #1 #2 #3 {{\sl Phys.\ Rev.\ B} {\bf #1}, #2 (#3)}
\def\prc #1 #2 #3 {{\sl Phys.\ Rev.\C} {\bf #1}, #2 (#3)}
\def\prd #1 #2 #3 {{\sl Phys.\ Rev.\ D} {\bf #1}, #2 (#3)}
\def\pre #1 #2 #3 {{\sl Phys.\ Rev.\ E} {\bf #1}, #2 (#3)}
\def\prept #1 #2 #3 {{\sl Phys.\ Repts.} {\bf #1}, #2 (#3)}
\def\prk #1 #2 #3 {{\sl Prog.\ Reac.\ Kinetics} {\bf #1}, #2 (#3)}
\def\prl #1 #2 #3 {{\sl Phys.\ Rev.\ Lett.} {\bf #1}, #2 (#3)}
\def\prsl #1 #2 #3 {{\sl Proc.\ Roy.\ Soc.\ London Ser. A} {\bf #1}, #2 (#3)}
\def\pss #1 #2 #3 {{\sl Prog.\ Surf.\ Sci.} {\bf #1}, #2 (#3)}
\def\pt #1 #2 #3 {{\sl Phys.\ Today} {\bf #1}, #2 (#3)}
\def\ptech #1 #2 #3 {{\sl Powder Tech.} {\bf #1}, #2 (#3)}
\def\ptrs #1 #2 #3 {{\sl Phil.\ Trans.\ Roy.\ Soc., Ser. A} {\bf #1}, #2 (#3)}
\def\rjpc #1 #2 #3 {{\sl Russ.\ J. Phys.\ Chem.} {\bf #1}, #2 (#3)}
\def\rmp #1 #2 #3 {{\sl Rev.\ Mod.\ Phys.} {\bf #1}, #2 (#3)}
\def\rpp #1 #2 #3 {{\sl Rep.\ Prog.\ Phys.} {\bf #1}, #2 (#3)}
\def\sci #1 #2 #3 {{\sl Science} {\bf #1}, #2 (#3)}
\def\sciam #1 #2 #3 {{\sl Scientific American} {\bf #1}, #2 (#3)}
\def\SIAMjam #1 #2 #3 {{\sl SIAM J.\ Appl.\ Math.} {\bf #1}, #2 (#3)}
\def\spu #1 #2 #3 {{\sl Sov.\ Phys.\ Usp.} {\bf #1}, #2 (#3)}
\def\SPEre #1 #2 #3 {{\sl SPE Reservoir Eng.} {\bf #1}, #2 (#3)}
\def\ssc #1 #2 #3 {{\sl Sol.\ State.\ Commun.} {\bf #1}, #2 (#3)}
\def\ss #1 #2 #3 {{\sl Surf.\ Sci.} {\bf #1}, #2 (#3)}
\def\stec #1 #2 #3 {{\sl Sov.\ Theor.\ Exp.\ Chem.} {\bf #1}, #2 (#3)}
\def\tAIME #1 #2 #3 {{\sl Trans.\ AIME} {\bf #1}, #2 (#3)}
\def\tfs #1 #2 #3 {{\sl Trans.\ Faraday Soc.} {\bf #1}, #2 (#3)}
\def\tpa #1 #2 #3 {{\sl Theor.\ Prob.\ Appl.} {\bf #1}, #2 (#3)}
\def\tpm #1 #2 #3 {{\sl Transport in Porous Media} {\bf #1}, #2 (#3)}
\def\usp #1 #2 #3 {{\sl Sov.\ Phys. -- Usp.} {\bf #1}, #2 (#3)}
\def\wrr #1 #2 #3 {{\sl Water Resources Res.} {\bf #1}, #2 (#3)}
\def\zpb #1 #2 #3 {{\sl Z. Phys.\ B} {\bf #1}, #2 (#3)}
\def\zpc #1 #2 #3 {{\sl Z. Phys.\ Chem.} {\bf #1}, #2 (#3)}
\def\ztf #1 #2 #3 {{\sl Zh. Tekh.\ Fiz.} {\bf #1}, #2 (#3)}
\def\zw #1 #2 #3 {{\sl Z. Wahrsch.\ verw.\ Gebiete} {\bf #1}, #2 (#3)}

\centerline{\bf Kinetics of a Diffusive Capture Process:  
Lamb Besieged by a Pride of Lions}
\vskip 0.4in
\centerline{P.~L.~Krapivsky{$^1$} and S.~Redner{$^2$}}
\bigskip 
\centerline{$^1$\sl Courant Institute of Mathematical Sciences}
\centerline{\sl New York University, New York, NY 10012-1185}
\medskip
\centerline{$^2$\sl Center for Polymer Studies and Department of Physics}
\centerline{\sl Boston University, Boston, MA 02215}

\vskip 0.3in

{
\narrower\narrower

\noindent The survival probability, $S_N(t)$, of a diffusing prey
(``lamb'') in the proximity of $N$ diffusing predators (a ``pride of
lions'') in one dimension is investigated.  When the lions are all to
one side of the lamb, the survival probability decays as a non-universal
power law, $S_N(t)\propto t^{-\b_N}$, with the decay exponent $\b_N$
proportional to $\ln N$.  The crossover behavior as a function of the
relative diffusivities of the lions and the lamb is also discussed.  When
$N\to\infty$, the lamb survival probability exhibits a log-normal decay,
$S_\infty(t)\propto \exp(-\ln^2 t)$.

}

\vskip 0.6in

\rchap{INTRODUCTION}

Consider a particle system which consists of a diffusing prey and $N$
independent, diffusing predators, with $N$ either finite or infinite.
The prey is absorbed, or dies, whenever it is touched by any of the
predators.  We are interested in the probability $S_N(t)$ for this
``lamb'' to survive until time $t$ when it is besieged by these
predatory ``lions'' [1].  While this appears to be a simple problem,
there are important aspects of the long-time behavior which are
incompletely understood.  Their resolution has fundamental
ramifications for diffusive processes in the presence of complex
absorbing boundaries and also practical implications, as this type of
capture process appears in a variety of applications, such as
diffusion-controlled chemical kinetics, wetting, melting, and
commensurate-incommensurate transitions (see, \eg, Refs.~[2-4]).  It
is known rigorously that for spatial dimension $d\geq 3$, the capture
process is ``unsuccessful'' (in the terminology of Ref.~1), in that
there is a finite probability for the lamb the survive indefinitely
for any $N$ and for any initial spatial distribution of the lions.
This result is a consequence of the transience of random walks for
$d>2$ [5].  For $d=2$, the capture process is ``successful'' -- the
lamb dies with probability one.  However, diffusing lions in two
dimensions are still sufficiently poor predators that the average
lifetime of the lamb is infinite.  Moreover, $S_N(t)\propto
[S_1(t)]^N\propto [\ln t]^{-N}$, \ie, the many-body nature of the
capture process is basically irrelevant. 

In one dimension, however, diffusing lions are more efficient in their
predation because of the recurrence of random walks [5].  This typically
leads to a lamb survival probability which decays as a power law in
time.  More generally, we may consider the survival probability,
$S_{N,M}(t)= S_{M,N}(t)$, when initially $N$ lions are placed to the
left and $M$ lions are placed to the right of the lamb.  The most
interesting situation is when initially $N$ lions are to one side of the
lamb.  For this initial condition, realizations in which the lamb runs
away from the lions leads to an anomalously slow decay of the lamb
survival probability.  Our primary result is to determine the asymptotic
behavior of the lamb survival probability in this predatory environment,
$S_{N,0}(t)\equiv S_N(t)$.  For finite $N$, we predict that $S_N(t)\sim
t^{-\b_N}$ with $\b_N\propto\ln N$, while $S_\infty(t)\propto
\exp(-\ln^2t)$ for $N\to\infty$.  We shall argue that these dependences
arise from the fact that the motion of ``closest'' lion (whose
individual identity can change with time) is enhanced compared to normal
diffusion.  When the number of lions is finite, this enhancement
manifests itself in the diffusivity of the last lion being proportional
to $\ln N$.  This factor is ultimately responsible for the logarithmic
increase of $\b_N$ on $N$.  In the limit $N\to\infty$, the co-ordinate
of the last lion actually varies as $\sqrt{t\ln t}$, and leads to the
lamb survival probability varying as $\exp(-\ln^2t)$.

To provide context for our results, consider first a system consisting
of one lamb and one lion, \ie, $(M,N)=(1,0)$ or $(0,1)$.  The survival
probability is trivially calculable in this case, since the distance
between the lion and the lamb undergoes pure diffusion with an
associated diffusivity $D_L+D_\ell$.  Here $D_L$ and $D_\ell$ are the
lion and lamb diffusivities, respectively.  Because of the equivalence
to diffusion, the survival probability is [5]
$$
S_{1,0}(t)\sim{x_0\over\sqrt{(D_L+D_\ell)t}},\eqnoi
$$ 
where $x_0$ is the initial separation between the lion and the lamb.
A more interesting situation is that of two lions with either: (i) one
lion on either side of the lamb (``trapped'' lamb), or (ii) both lions
to one side (``chased'' lamb).  These two systems can be
straightforwardly solved when the diffusivities of all three particles
are different.  For the trapped lamb, denote the particle positions as
$x_1$, $x_2$, and $x_3$, with 1 and 3 referring to the surrounding
lions, and 2 to the trapped lamb.  Let the corresponding diffusivities
be $D_1$, $D_2$, and $D_3$.  To solve for the survival probability, it
is convenient to introduce the rescaled co-ordinates
$y_i=x_i/\sqrt{D_i}$, each of which diffuses at the same rate.  The
survival of the lamb corresponds to the constraints
$y_1\sqrt{D_1}<y_2\sqrt{D_2}$ and $y_2\sqrt{D_2}<y_3\sqrt{D_3}$.
Since the co-ordinates $y_i$ diffuse isotropically, lamb survival is
equivalent to the survival of a random walk in three-dimensional space
within the wedge-shaped region bounded by the absorbing planes
$y_1\sqrt{D_1}=y_2\sqrt{D_2}$ and $y_2\sqrt{D_2}=y_3\sqrt{D_3}$.  By
straightforward geometric considerations [2], this three-dimensional
diffusion process is equivalent to diffusion in two dimensions within
a wedge of opening angle

$$
\theta= 
\cos^{-1}\left({D_2\over\sqrt{(D_1+D_2)(D_2+D_3)}}\right)
\eqnoi
$$
For this two-dimensional problem, it is well-known that the survival
probability asymptotically varies as $t^{-\pi/2\theta}$\ref{6}.
Identifying $D_1=D_3=D_L$ and $D_2=D_\ell$, leads to
$$
\b_{1,1}(\CR)=
\left[{2\over\pi}\cos^{-1}{\CR\over{1+\CR}}\right]^{-1},
\eqnai
$$
where $\CR\equiv D_\ell/D_L$.  Similarly, one finds for
the ``chased'' lamb
$$
\b_{2,0}=\left[2-{2\over\pi}\cos^{-1}{\CR\over{1+\CR}}\right]^{-1}
\eqnbi
$$

Physically, when $\CR\to\infty$, the motion of the lions becomes
irrelevant and the trapped lamb problem maps to a diffusing particle
in a fixed size absorbing domain.  For this geometry, the survival
probability decays exponentially in time, corresponding to
$\b_{1,1}\to\infty$ in \lasta.  Conversely, for the lamb at rest,
$\CR\to 0$, the survival probability is the square of the
corresponding survival probability in the two-particle system so that
$\b_{1,1}\to 1$.  For the chased lamb (\lastb), $\b_{2,0}$ has the
limiting values $\b_{2,0}= 1/2$ for $\CR\to\infty$ and $\b_{2,0}= 1$
for $\CR\to 0$.  These two values are in accord with a direct
consideration of the extreme cases of immobile lions or immobile lamb,
respectively.  Notice also that for $\CR=1$ we get $\b_{1,1}=3/2$ and
$\b_{2,0}=3/4$, results which were known previously [1].

Extending the above approach, the survival of a lamb in the presence
of $N>2$ lions can be mapped onto the survival of an $N+1$-dimensional
random walk which is confined within an absorbing hyper-wedge.  This
problem does not appear to be generally soluble, however.
Numerically, there has been investigation\ref{1} of the one-sided
equal-diffusivities problem for the cases $N=3$, 4, and 10 as part of
an effort to understand the general behavior on $N$.  This simulation
reveals that the exponent $\beta_{N}$ grows slowly with $N$, with
$\b_3= 0.91$, $\b_4\approx 1.032\pm 0.01$, and $\b_{10}\approx 1.4$.
(Because the case $N=4$ is close to the transition between a finite
and infinite lamb lifetime, there is more numerical data and hence a
greater precision in the estimate for $\b_4$.)~ The understanding of
this slow dependence of $\b_N$ on $N$ is the focus of our work.

In the next section, we provide a heuristic argument for the dependence
of $\b_N$ on $N$, as well the behavior for $N\to\infty$.  A more
complete derivation of these results is given Sec.\ III.  The general
dependence of the survival probability on the ratio $\CR=D_\ell/D_L$ is
also considered.  In Sec.\ IV, we treat the case of $N\to\infty$.  A
general discussion and conclusions are given in Sec.\ V.

\rchap{HEURISTIC ARGUMENTS FOR ONE-SIDED SIEGE}

First consider the trivial case of a stationary lamb, $D_\ell=0$.  For
non-interacting lions, the lamb survival probability is just the product
of the survival probabilities associated with each lion-lamb pair.  This
immediately gives $S_N(t)\propto (t^{-1/2})^N$, from which
$\b_N(0)=N/2$.  For this case, the relative positions of the lions do
not matter in the asymptotic behavior of $S_N(t)$, \ie, it is immaterial
whether the lamb is ``trapped'' or ``chased''.  When the lamb also
diffuses, it is convenient to consider the survival probability in the
rest frame of the lamb.  Although the lions still diffuse independently,
their relative motions with respect to the lamb are not independent.
Therefore to determine the survival probability of a diffusing lamb, it
more useful to track the position of the closest lion only.  For
concreteness and simplicity, suppose initially that all the lions are at
the origin and the lamb is at $x_0>0$.  A rough estimate for the
location of the closest or ``last'' lion, $x_+(t)$, is provided by
$$
\int_{x_+(t)}^\infty {1\over\sqrt{4\pi D_Lt}}\,e^{-x^2/4D_Lt}\,\,dx= 1/N.
\eqnoi
$$
This specifies that there should be one lion in the range
$(x_+(t),\infty)$ out of an initial group of $N$ lions.
By asymptotic expansion of this integral, the location of the last lion
is given by
$$
x_+(t)\sim\sqrt{4D_Lt\,\ln N}.
\eqnai
$$
In the limit $N\to\infty$, a physically tractable initial condition is
to have a uniform density of lions $c_0$ extending from $-\infty$ to 0.
In this situation, only a number $N\propto\sqrt{c^2_0 D_L t}$ 
of the lions are ``dangerous'', \ie, are 
potential candidates for being closest to the lamb.  Consequently,
for $N\to\infty$, the leading behavior of $x_+(t)$ is
$$
x_+(t)\sim \sqrt{2D_Lt\,\ln(c^2_0 D_L t)}\qquad N\to\infty.
\eqnbi
$$

The next step in our heuristic approach is to posit that for large $N$,
the true stochastic motion of the last lion can be replaced by a
continuous motion $x(t)$ with $x(t)=x_+(t)$, as given in \lasts.  Then
the system reduces to a two-body problem of a diffusing lamb and an
approaching absorbing boundary, whose location is $x_+(t)=\sqrt{At}$.
As discussed in the next section, the survival probability of a lamb
adjacent to such an approaching ``cliff'' can be calculated
analytically.  This gives the exponent of the survival probability as
$\b\sim A/16D_\ell$.  Substituting the appropriate value of $A$ as
specified by \lasts, we obtain $S_N(t)\sim t^{-\b_N(\CR)}$ with
$\b_N(\CR)\sim \ln(N\CR)/4\CR$ for finite $N$, and
$S_{\infty}(t)\sim\exp(-\ln^2t)$ for $N\to\infty$.

\rchap{ASYMPTOTIC ANALYSIS FOR ONE-SIDED SIEGE}

A more rigorous approach is to consider the survival probability in an
auxiliary ``deadline'' problem whose asymptotic behavior turns out to
give a tight lower bound for the true survival probability of the
lamb.  The deadline problem is defined as follows: Consider an
imaginary point $x_{\rm dead}(t)$ between the lamb and the lions which
moves deterministically according to $x_{\rm dead}(t)=\sqrt{At}$.  If
the lamb crosses this line, it is considered to have died;
analogously, if any of the lions overtakes the deadline, the lamb is
again considered to have died.  Our strategy is to determine the
survival probability in this auxiliary problem, and then maximize this
probability with respect to the parameter $A$.  First notice that a
deadline position which is proportional to $\sqrt{t}$ already
optimizes the lamb survival probability with respect to other
power-law motions for the deadline.  That is, is $x_{\rm dead}(t)$
were proportional to $t^\a$ with $\a<1/2$, we would asymptotically
recover the behavior for the stationary deadline, which grossly
overestimates the decay exponent as $\b_N=(N+1)/2$.  Conversely, for
faster than diffusive deadline motion, \ie, $\a>1/2$, the probability
that the lamb does not hit the deadline decays as a stretched
exponential [7]; therefore, this case can be also ignored.  The
marginal situation of $\a=1/2$ thus plays a fundamental role.

To compute the survival probability for the deadline problem, we have to
solve two first-passage problems: (i) The survival of a diffusing
particle in the proximity of a {\it receding} absorbing boundary, or
cliff.  This corresponds to a single lion, and we define the probability
that a lion does not reach the cliff to be $S_{\rm lion}(t)\propto
t^{-\b_{\rm lion}(D_L,A)}$. (ii) The survival of a diffusing particle in
the proximity of an approaching cliff.  This corresponds to the lamb,
and the associated survival probability is defined as $S_{\rm
lamb}(t)\propto t^{-\b_{\rm lamb}(D_\ell,A)}$.  The full survival
probability $S_N(t)\propto t^{-\b_N}$ is clearly the product $S_{\rm
lion}^N(t)\,S_{\rm lamb}(t)$, so that $\b_N=\b_{\rm
lamb}(D_\ell,A)+N\b_{\rm lion}(D_L,A)$.  Once we know the exponents
$\b_{\rm lion}$ and $\b_{\rm lamb}$, we optimize the decay exponent
$\b_N$ with respect to the amplitude $A$.  By appealing to the method of
the ``optimal fluctuation'' [8], we hypothesize that this extremal
survival probability in the deadline problem gives the true asymptotic
behavior.

Fortunately, the exponents $\b_{\rm lamb}$ and $\b_{\rm lion}$ have been
computed in various physical [9] and mathematical studies [10] so that
the full deadline problem is soluble.  For completeness, however, we
outline our approach, given in Ref.~[7], which has the advantage of
conceptual and technical simplicity.  While this earlier work considered
only the case of a receding cliff (relevant for the lions), the
extension to the case of an approaching cliff can be derived with
minimal additional effort.  Let us therefore recall the steps in the
computation of the survival probability for the case of the receding
cliff.  Consider a lion which is initially placed on the negative
$x$-axis and that the cliff position is $x_0(t)=\sqrt{At}$.  In the
long-time limit, the lion density approaches the scaling form [7]

$$
c(x,t) \sim t^{-\b_{\rm lion}-1/2}\CC(\xi),
\eqnoi
$$ 
where $\xi=1-{x\over x_0}$ is the appropriate dimensionless distance
variable and\ $\CC(\xi)$ is a scaling function.  The initial co-ordinate
of the lion, $-\infty<x\leq x_0$, corresponds to $0\leq \xi<\infty$.
The power law prefactor is chosen to ensure that the survival
probability decays as $t^{-\b_{\rm lion}}$, as defined previously.

Substituting \last\ into the diffusion equation, one finds that the
scaling function satisfies
$$
{D_L\over A}{d^2\CC \over d\xi^2} +{1\over 2}(\xi-1)\td\CC \xi+
\left(\b_{\rm lion}+{1\over 2}\right)\CC=0.
\eqnoi
$$
Introducing the transformation
$$
\xi-1=\sqrt{2D_L\over A}~\eta, 
\quad \CC(\xi)=\exp\left(-{\eta^2\over 4}\right)\CD(\eta),
\eqnoi
$$
one finds that $\CD(\eta)$ satisfies the parabolic cylinder equation of
order $2\b_{\rm lion}$ [12],
$$
{d^2\CD_{2\b_{\rm lion}}\over d\eta^2}+
\left[2\b_{\rm lion}+{1\over 2}-{\eta^2\over 4}\right]\CD_{2\b_{\rm lion}}=0.
\eqnoi
$$
The absorbing boundary condition at the edge of the
cliff implies
$$
\CD\left(-\sqrt{A/2D_L}\right)=0.
\eqnai
$$
On the other hand, to avoid a singular solution at $\eta=\infty$,
the second boundary condition is
$$
\CD(\eta=\infty)=0.
\eqnbi
$$

Mathematically, the determination of $\b_{\rm lion}$ and $\CD(\eta)$ is
equivalent to finding the ground state energy and wave function of a
quantum particle in a potential composed of an infinite barrier at
$\eta=-\sqrt{A/2D_L}$ and the harmonic oscillator potential for
$\eta>-\sqrt{A/2D_L}$ [11].  Higher excited states do not contribute in the
long time limit.  This relation with quantum mechanics allows one to
apply well-known techniques to determine the asymptotic behavior [7,10].
Among the two elemental solutions of the parabolic cylinder equation,
$\CD_{2\b_{\rm lion}}(\eta)$ and $\CD_{2\b_{\rm lion}}(-\eta)$, only the
former satisfies the boundary condition $\CD(\infty)=0$.  Therefore, the
absorbing boundary condition of \lasta\ determines the decay exponent
$\b_{\rm lion}=\b_{\rm lion}(D_L,A)$.  

As discussed previously, the interesting behavior emerges in the large
$N$ limit.  For this case, the deadline position grows as $\sqrt{At}$
but with an anomalously large amplitude $A$.  Consequently, the
probability distribution of each lion is only weakly affected by the
receding deadline.  This allows us to employ the ``free particle''
particle Gaussian approximation for the probability distribution of each
lion.  Although this form does not satisfy the absorbing boundary
condition, the error is negligible because $A\gg 1$.  Consequently, we
can determine the decay exponent simply by computing the flux to the
absorbing boundary for the assumed Gaussian probability distribution
[7].  This yields
$$
\b_{\rm lion}(D_L,A)\simeq \sqrt{A\over 4\pi D_L}\,e^{-A/4D_L}.
\eqnoi
$$
In the limit $A\to\infty$, this simple-minded approach coincides
with the results from a complete analysis in terms of the
parabolic cylinder function solution.

An analogous, but simpler, treatment applies for the approaching cliff,
which we use to describe the interaction of the deadline with
the lamb.  That is, suppose that a lamb is initially placed on the
positive $x$-axis and that there is an approaching cliff whose location
is at $\sqrt{At}$.  To solve this problem by the same approach as the
receding cliff, we introduce the appropriate dimensionless length variable 
$\xi={x\over x_0}-1$ and  make the analogous scaling ansatz as in
\back5, so that \back4\ is replaced by
$$
{D_\ell\over A}{d^2\CC \over d\xi^2} +{1\over 2}(\xi+1)\td\CC \xi+
\left(\b_{\rm lamb}+{1\over 2}\right)\CC=0.
\eqnoi
$$
For this case, it is again helpful to introduce $\eta$ via
$\xi+1=\sqrt{2D_\ell\over A}~\eta$ and
$\CC(\xi)=e^{-\eta^2/4}\,\CD(\eta)$.  The scaling function $\CD(\eta)$
is again the parabolic cylinder function of order $2\b_{\rm lamb}$ and
the absorbing boundary condition,
$$
\CD_{2\b_{\rm lamb}}\bigl(\sqrt{A/2D_\ell}\bigr)=0,
\eqnoi
$$ 
now determines the decay exponent $\b_{\rm lamb}=\b_{\rm
lamb}(D_\ell,A)$.  Since the relevant zero of the parabolic cylinder
function $\eta=\sqrt{A/2D_\ell}$ is large, $\b_{\rm lamb}$ is also
large.  Then an inspection of \back4\ provides the estimate $2\b_{\rm
lamb}+{1\over 2}\simeq {\eta^2\over 4}$, or
$$
\b_{\rm lamb}(D_\ell,A)\simeq {A\over 16D_\ell}.
\eqnoi
$$
Therefore the total decay exponent for the deadline problem is
$$
\b_N(\CR,A)=\b_{\rm lamb}(D_\ell,A)+N\b_{\rm lion}(D_L,A)
\simeq {A\over 16D_\ell}+N\sqrt{A\over 4\pi D_L}\,e^{-A/4D_L}.
\eqnoi
$$
Minimizing this expression with respect to $A$ yields the optimal value
$A^\ast\sim 4D_L\ln(4N\CR)$.  Thus the deadline motion is enhanced by a
factor of $\ln N$ compared simple diffusion; notice that this coincides
with the motion of the last lion in a pride of $N$ lions.
Correspondingly, the optimal value of the decay exponent 
$\b_N(\CR)\equiv \b_N(\CR,A^\ast)$ is
$$
\b_N(\CR)\sim {\ln(4N\CR)\over 4\CR}.
\eqnoi
$$

Our construction of the deadline problem relies on the assumption that
$N\gg 1$.  This assumption is crucial, otherwise the deadline problem
would not provide a meaningful approximation for the behavior of the
original system.  However, the physical nature of the problem suggests
that different asymptotic behaviors for the lamb survival probability
should arise for $\CR\gg 1$ and $\CR\ll 1$.  In fact, consideration of
the limiting cases of a stationary lamb and of stationary lions,
suggests that \last\ is actually valid only for $N^{-1}\ll \CR\ll \ln
N$.  In the slow-lamb limit, $\CR\ll N^{-1}$, the logarithmic behavior
of \last\ should cross over to that of the stationary-lamb limit,
namely, $\b_N(0)=N/2$.  In the complementary fast-lamb limit, $\CR\gg
\ln N$, the behavior of the stationary lion case should be recovered, in
which $\b_N(\infty)=1/2$.  Thus the full dependence of $\b_N$ on the
diffusivity ratio $\CR$ is expected to be

$$
\b_N(\CR)=\cases{N/2  &\qquad $\CR\ll 1/N$;  \cr
                      \cr
              \ln(4N\CR)/4\CR &\qquad $1/N\ll \CR\ll \ln N$; \cr
                      \cr
              1/2 &\qquad $\CR\gg \ln N$. \cr}
\eqnoi
$$ 
The non-universal dependence of $\b_N$ on the diffusivity ratio for the
intermediate regime of $1/N\ll \CR\ll \ln N$ is the generalization of
the exponents in Eqs.~(3), for the three-particle system, to arbitrary
$N$.

\rchap{INFINITE NUMBER OF LIONS}

Consider now a lamb which is under one-sided siege by an infinite pride
of lions.  (These lions need to be distributed over an infinite domain
so that their density is everywhere finite.  If the lion density were
infinite at some point, then the closest lion would move inexorably
toward to the lamb at each step, leading to the survival probability
decaying exponentially in time.)~ The interesting situation is when the
lions are all to one side of the lamb.  However, to introduce our
approach, it is instructive to consider first the simpler two-sided
problem, in which lions are uniformly and symmetrically distributed with
unit density on either side of a stationary lamb, a problem has been
previously investigated by asymptotic and exact methods [13,14].  For
completeness, we describe an approach which is in the spirit of the
previous section.

For a lamb at the origin, the density of the lions $c(x,t)$ may be found
by solving the diffusion equation with an absorbing boundary condition
at $x=0$ and with the initial condition of a unit density everywhere.
This yields [5]

$$
c_{\rm lamb}(x,t)={2\over \sqrt{\pi}}
\int_0^{|x|/\sqrt{4D_Lt}}d\zeta e^{-\zeta^2}.
\eqnoi
$$ 
Thus the diffusive flux of lions toward the lamb is 
$D_L\bigl(\pd c x\big|_{x=0^+}-\pd c x\big|_{x=0^-}\bigr)
=\sqrt{4D_L\over \pi t}$.  The survival probability $S_\infty(t)$
therefore obeys
$$
\td {S_\infty(t)} t=-S_\infty(t)\sqrt{4D_L\over \pi t},
\eqnoi
$$ 
with solution
$$
S_\infty(t)=\exp\left[-4\sqrt{D_Lt\over \pi}\,\,\right].
\eqnoi
$$ 
When both the lion and lamb are diffusing, a faster decay occurs.
However, since the dominant annihilation mechanism arises from the
diffusive flux of lions toward the lamb, we expect that the asymptotic
decay is still given by \last\ [13].  The crucial feature of this
two-sided problem is that there is no good ``survival'' strategy, so
that the lamb survival probability must decay rapidly in time.

For a lamb under one-sided siege, we again attempt a solution via the
auxiliary deadline model.  Assume that the deadline undergoes enhanced
square-root motion, \ie, $x_{\rm dead}(t)=\sqrt{At}$, with $A\gg 1$.
Repeating the steps employed previously for a finite pride of lions, we
have $S_\infty(t)=S_{\rm lion}(t)S_{\rm lamb}(t)$, with $S_{\rm
lamb}(t)\propto t^{-A/16D_\ell}$, as in the case of a finite pride.  To
determine $S_{\rm lion}(t)$ we again use a free particle approximation,
since we expect that the amplitude $A$ will be large.  Thus for the
probability density of the lions, we ignore the adsorbing boundary
condition on the moving deadline.  For the initial condition of unit
density of lions for $x<0$ and zero density otherwise, the time
dependent lion density is [5]

$$
c_{\rm lion}(x,t)={1\over \sqrt{\pi}}
\int_{x/\sqrt{4D_Lt}}^\infty \,\,d\zeta\,\, e^{-\zeta^2}.
\eqnoi
$$ 
Although this solution disagrees with the adsorbing boundary condition
on the deadline, the disagreement is of order $e^{-A/4D_L}$ and is
negligible when $A\gg 1$.

Computing the diffusive flux of lions through the deadline, we make use of
\last\ and $A/D_L\gg 1$ to find 
$$
-D_L\pd cx\Big|_{x=x_{\rm dead}(t)}\simeq \sqrt{D_L\over{4\pi t}}e^{-A/4D_L}.
\eqnoi
$$ 
The lion survival probability, $S_{\rm lion}(t)$, therefore obeys
$$
\td {S_{\rm lion}(t)} t\simeq 
-S_{\rm lion}(t)\sqrt{D_L\over{4\pi t}}e^{-A/4D_L},
\eqnoi
$$ 
with solution
$$
S_{\rm lion}(t)\simeq \exp\left[-e^{-A/4D_L}\sqrt{D_Lt\over \pi}\,\,\right].
\eqnoi
$$ 
Thus the full survival probability is
$$
S_\infty(t)\propto \exp\left[-e^{-A/4D_L}\sqrt{D_Lt\over \pi}
-{A\over 16D_\ell}\ln t\right].
\eqnoi
$$ 

Maximizing this survival probability with respect to $A$, we find 
that optimal value, $A^\ast$,  grows in time as,
$$
A^\ast\sim 2D_L\ln\left[{t\over \ln^2 t}\right].
\eqnoi
$$ 
The leading logarithmic behavior is in accord with our naive estimate
given in Sec.\ II.  Combining \backs1\ and \lastn\ gives
$$
S_\infty(t)\propto \exp\left[-\ln^2 t\right].
\eqnoi
$$ 
Thus we obtain a survival probability for the one-sided system which
decays {\it faster} than any power law and {\it slower} than any
stretched exponential.  The decay is, however, universal in that the
power of logarithm does {\it not} depend on the diffusivity ratio
$\CR=D_\ell/D_L$.

\rchap{SUMMARY AND DISCUSSION}

For a diffusing lamb in one dimension which is adjacent to a pride of
$N$ diffusing, predatory lions, the survival probability of the lamb
decays as $S_N(t)\sim t^{-\b_N}$ with $\b_N$ proportional to $\ln N$.
This slow increase of $\b_N$ on $N$ reflects the fact that the dominant
contribution to the survival probability arises from realizations in
which the lamb ``runs away'' from the lions.  Consequently, each
additional lion in the system has a progressively weaker effect on the
lamb survival.  This is in contrast to the case of a stationary lamb,
where each additional lion is equally effective in hunting the lamb, so
that $\b_N$ is proportional to $N$.  The exponent $\b_N$ is also a
decreasing function of the diffusivity ratio, $\CR=D_\ell/D_L$, with
$\b_N=N/2$ for $\CR=0$ and $\b_N=1/2$ for $\CR=\infty$.  Thus, in accord
with intuition, the best survival strategy for the lamb is to diffuse
faster than the lions.  In contrast, for a two-sided system, where the
lions initially surround the lamb, the best survival strategy for the
lamb is to remain still.

The above non-universal power-law decay of $S_N(t)$ motivated the basic
question, considered in Ref.~1, of whether the mean lamb lifetime
$$
\t_N\equiv -\int_0^\infty dt\,\,t\,\td{S_N(t)}{t}
    =\int_0^\infty dt\,S_N(t)
\eqnoi
$$ 
is finite or infinite.  From \last, it is clear that $\t_N$ is finite
for $\b_N>1$, and $\t_N$ diverges otherwise.  The numerical evidence
from Ref.~1 indicates that when $\CR=1$ the lamb lifetime is finite
for $N\geq 4$.  Since our prediction for $\b_N$ is anticipated to be
accurate only for large $N$, we may conclude that the lamb lifetime is
finite when $N\geq N^{\ast}(\CR)$, but cannot provide an accurate
estimate of this threshold value.  Additionally, we predict that
$N^\ast(\CR)$ should increase rapidly with the diffusivity ratio
$D_\ell/D_L$, namely $\ln N^\ast(\CR)\propto \CR$ (Eq.~(17)).

For an infinite number of predators, the lamb survival probability
$S_\infty(t)$ exhibits a log-normal decay $\exp(-\ln^2t)$.  This
contrasts sharply with the corresponding behavior in the two-sided
geometry, where $S_\infty(t)\propto \exp(-t^{1/2})$.  For the one-sided
geometry, it is striking that this same survival probability occurs for
a reactive system consisting of a single ``fast impurity'' which moves
with velocity $v>1$ within a semi-infinite sea of mutually annihilating
ballistic particles moving at velocity $v=\pm 1$ [15].  Given the
superficial similarity of the fast impurity with the lions and lamb
systems, it may be interesting to seek a deeper connection between these
two problems.

We close with mention of a generalization where the lions are
``vicious'' among themselves, in addition to stalking the lamb.  (For
such self-predatory lions, their number must be infinite; otherwise, the
lamb survival probability has a non-zero asymptotic value.)~ There are
two natural possibilities for the outcome when two lions meet: either
(i) one lion dies (aggregation), or (ii) both die (annihilation).  The
first possibility is particularly simple, since the closest lion
undergoes pure diffusion, independent of its individual identity.  Thus
the two-sided geometry reduces to the finite particle system
$(M,N)=(1,1)$, with decay exponent given by Eq.~(3a).  The one-sided
aggregation problem is even simpler since it reduces to the (1,0)
problem whose solution is given in Eq.~(1).

Annihilating lions framework leads to more interesting behavior, as the
position of the closest lion suddenly jumps away from the lamb whenever
the closest two lions annihilate.  The two-sided version of this problem
was introduced in Ref.~16.  It was found that the lamb survival
probability decays as a non-universal power law, $S(t)\propto
t^{-\g(\CR)}$, with a Smoluchowski theory predicting
$\g(\CR)=\sqrt{(1+\CR)/8}$.  This agrees with the obvious exact result
$\g(1)=1/2$ and is close to the exact value $\g(0)=3/8$ [17].  This
Smoluchowski prediction and also provides a good approximation for the
simulation results for arbitrary $\CR$ [16].  Generalizations of the
two-sided annihilation problems (\eg, to many dimensions) have also been
discussed in [16-18].  To the best of our knowledge, however, lamb
survival in the presence of a one-sided distribution of annihilating
lions has not yet been treated.  If one naively assumes that the
two-sided death probability can be expressed in terms of independent
one-sided death probabilities, then the exponents of the one-sided
system, $\b(\CR)$, and the two-sided system, $\g(\CR)$, are simply
related by $\b(\CR)=\g(\CR)/2$.  This is clearly correct for $\CR=0$,
where the independence of the one-sided death probabilities is exact.
Consequently, the known value of $\g(0)$ gives $\b(0)=3/16$.  However
for $\CR>0$, the motions of the lions are not independent in the rest
frame of the lamb, and the independence of one-sided killing
probabilities is only an approximation.  Interestingly, however,
simulations suggest that for equal lion and lamb diffusivities,
$\b(\CR=1)=1/4$, which conforms to the relation $\b(\CR)=\g(\CR)/2$.  We
do not have an understanding of this simple yet paradoxical result.
Finally, as $\CR\to\infty$, it is clear that $\b(\CR)\to 1/2$.  Thus we
conclude that $\b(\CR)$ is a slowly increasing function of $\CR$, with
$\b(0)=3/16$ and $\b(\infty)=1/2$.

\bigskip\bigskip
\centerline{\bf ACKNOWLEDGMENTS}
\medskip

We thank D. Griffeath for helpful correspondence.  We also gratefully
acknowledge NSF grant DMR-9219845 and ARO grant \#DAAH04-93-G-0021 for
partial support of this research.

\baselineskip=14truebp
\vskip 0.6 in
\centerline{\bf REFERENCES}
\bigskip

\refi M.~Bramson and D.~Griffeath, ``Capture problems for coupled
      random walks'', in {\it Random walks,  
      Brownian motion, and interacting particle
      systems: a festschrift in honor of Frank Spitzer}, eds. 
      R.~Durrett and H.~Kesten, (Boston, Birkhauser, 1991), 153-188.

\refi D.~ben-Avraham, ``Computer simulation methods for 
      diffusion-controlled reactions'', 
      {\sl J.\ Chem.\ Phys.} {\bf 88}, 941-947 (1988);
      M.~E.~Fisher and M.~P.~Gelfand, ``The reunions of three dissimilar
      vicious walkers'', {\sl J. Stat.\ Phys.} {\bf 53}, 175-189 (1988).

\refi M.~E.~Fisher, ``Walks, walls, wetting, and melting'', 
      {\sl J. Stat.\ Phys.} {\bf 34}, 667-729 (1984).

\refi D.~A.~Huse and M.~E.~Fisher, ``Commensurate melting, domain walls,
      and dislocations'',  {\sl Phys.\ Rev.\ B} {\bf 29}, 239-270 (1984);
      R. Lipowsky and T. M. Nieuwenhuizen, ``Intermediate fluctuation
      regime for wetting transitions in two dimensions'', 
      {\sl J. Phys.\ A} {\bf 21}, L89-L94 (1988);
      P.~J.~Forrester, ``Probability of survival for vicious walkers
      near a cliff'', {\sl J. Phys.\ A} {\bf 22}, L609-L613 (1989);
      J.~W.~Essam and A.~J.~Guttmann, ``Vicious walkers and directed
      polymer networks in general dimensions'', {\sl Phys.\ Rev.\ E} 
      {\bf 52}, 5849-5862 (1995). 

\refi G.~H.~Weiss and R.~J.~Rubin, ``Random walks: Theory and selected
      applications'', {\sl Adv.\ Chem.\ Phys.}  {\bf 52}, 363-505
      (1983), and references therein; see also N. G. van Kampen, {\it
      Stochastic Processes in Physics and Chemistry} (North-Holland,
      Amsterdam, 1981).
      
\refi H. S. Carslaw and J. C. Jaeger, {\sl Conduction of Heat in Solids}
      Chap.\ XI (Oxford University Press, Oxford, 1959); E.~B.~Dynkin
      and A.~A.~Yushkevich, {\it Markov Processes: Theorems and
      Problems} (Plenum Press, NY, 1969); D.~L.~Burkholder, ``Exit times
      of Brownian motion, harmonic majorization, and Hardy spaces'' 
      {\sl Adv. Math.} {\bf 26}, 182-205 (1977); R.~D.~De Blassie, 
      ``Exit times from cones in {\bf R}$^n$ of Brownian motion'', 
      {\sl Z. Wahr. verb. Gebiete} {\bf 74}, 1-29 (1987).
 
\refi P.~L.~Krapivsky and S.~Redner, ``Life and death in an expanding 
      cage and at the edge of a receding cliff'',  {\sl Amer.\ J.\ Phys.},
      to appear.

\refi I.~M.~Lifshits, S.~A.~Gredeskul, and L.~A.~Pastur,
      {\it Introduction to the theory of disordered systems}
      (Wiley, New York, 1988).

\refi L.~Turban, ``Anisotropic critical phenomena in parabolic 
      geometries: The directed self-avoiding walk'', {\sl J. Phys.\ A}
      {\bf 25}, L127-L134 (1992);  F.~Igl\'oi, ``Directed polymer 
      inside a parabola: Exact solution'', {\sl Phys.\ Rev.\ A} 
      {\bf 45}, 7024-7029 (1992);  F.~Igl\'oi, I.~Peschel, 
      and L.~Turban, ``Inhomogeneous systems with unusual critical 
      behaviour'', {\sl Adv.\ Phys.} {\bf 42}, 683-740 (1993).

\refi L.~Breiman, ``First exit time from the square root boundary'', 
      {\sl Proc.\ Fifth Berkeley Symp.\ Math.\ Statist.\ and Probab.}, 
      {\bf 2}, 9-16 (1966);
      H.~E.~Daniels, ``The minimum of a stationary Markov superimposed
      on a U-shape trend'', {\sl J.\ Appl.\ Prob.} {\bf 6}, 399-408 (1969); 
      K.~Uchiyama, ``Brownian first exit from sojourn over one sided
      moving boundary and application'', {\sl Z. Wahrsch.\ verw.\ Gebiete}
      {\bf 54}, 75-116 (1980);
      P.~Salminen, ``On the hitting time and the exit time for
      a Brownian motion to/from a moving boundary'', {\sl Adv.\ Appl.\ Prob.} 
      {\bf 20}, 411-426 (1988).

\refi L. I. Schiff, {\it Quantum Mechanics} (McGraw-Hill, New
      York, 1968).

\refi C.~M.~Bender and S.~A.~Orszag, {\it Advanced Mathematical Methods
      for Scientists and Engineers} (McGraw-Hill, New York, 1978).

\refi S.~Redner and K.~Kang, ``Kinetics of the scavenger reaction'',
      {\sl J. Phys.\ A} {\bf 17}, L451-L455 (1984).

\refi A.~Blumen, G.~Zumofen, and J.~Klafter, ``Target Annihilation by
      Random Walkers'', {\sl Phys.\ Rev.\ B} {\bf 30}, 5379-5382 (1984).

\refi P.~L.~Krapivsky, S.~Redner, and F.~Leyvraz,  ``Ballistic
      annihilation kinetics: The case of discrete velocity distributions'',
      {\sl Phys.\ Rev.\ E} {\bf 51}, 3977-3987 (1995). 

\refi P.~L.~Krapivsky, E.~Ben-Naim, and S.~Redner, 
      ``Kinetics of heterogeneous single-species annihilation'',
       {\sl Phys.\ Rev.\ E} {\bf 50}, 2474-2481 (1994).

\refi B.~Derrida, ``Exponents appearing in the zero-temperature dynamics
      of the 1D Potts model'', {\sl J. Phys.\ A} {\bf 28}, 1481-1491 (1995);
      B.~Derrida, V.~Hakim, and V.~Pasquier, ``Exact first-passage exponents
      of 1D domain growth: relation to a reaction-diffusion model'',
      {\sl Phys.\ Rev.\ Lett.} {\bf 75}, 751-754 (1995).

\refi J.~Cardy, ``Proportion of unaffected sites in a reaction-diffusion
      process'', {\sl J. Phys.\ A} {\bf 28}, L19-L24 (1995);
      E.~Ben-Naim, ``Reaction kinetics of cluster impurities'', 
      {\sl Phys.\ Rev.\ E} {\bf 53}, 1566-1571 (1996); 
      M.~Howard, ``Fluctuation Kinetics in a Multispecies Reaction-Diffusion 
      System'', {\it cond-mat 9510053}.

\vfill\eject\bye